# Mpox Screen Lite: AI-Driven On-Device Offline Mpox Screening for Low-Resource African Mpox Emergency Response

Yudara Kularathne[1], Prathapa Janitha[2], Sithira Ambepitiya[2],
[1] HeHealth Inc. San Francisco, CA, USA
[2] Aagee AI Pte. Ltd. Singapore

**Abstract**

Background: The 2024 Mpox outbreak, particularly severe in Africa with clade 1b emergence, has highlighted critical gaps in diagnostic capabilities in resource-limited settings. This study aimed to develop and validate an artificial intelligence (AI)-driven, on-device screening tool for Mpox, designed to function offline in low-resource environments.

Methods: We developed a YOLOv8n-based deep learning model trained on 2,700 images (900 each of Mpox, other skin conditions, and normal skin), including synthetic data. The model was validated on 360 images and tested on 540 images. A larger external validation was conducted using 1,500 independent images. Performance metrics included accuracy, precision, recall, F1-score, sensitivity, and specificity.

Findings: The model demonstrated high accuracy (96%) in the final test set. For Mpox detection, it achieved 93% precision, 97% recall, and an F1-score of 95%. Sensitivity and specificity for Mpox detection were 97% and 96%, respectively. Performance remained consistent in the larger external validation, confirming the model's robustness and generalizability.

Interpretation: This AI-driven screening tool offers a rapid, accurate, and scalable solution for Mpox detection in resource-constrained settings. Its offline functionality and high performance across diverse datasets suggest significant potential for improving Mpox surveillance and management, particularly in areas lacking traditional diagnostic infrastructure.

**Disclosures:** All authors are currently or were previously employed by HeHealth Inc.

**Acknowledgements and Funding:** A startup, HeHealth Inc has raised funding from institutional investors and angel investors

**Author contact information:** Dr. Yudara Kularathne, Email: yu@hehealth.ai

**Introduction**

The recent declaration of Mpox as a public health emergency of international concern (PHEIC) by the World Health Organization (WHO) in August 2024 underscores the urgent global health challenge posed by this evolving zoonotic disease.[1] Concurrently, the Africa Centres for Disease Control and Prevention (Africa CDC) elevated Mpox to a public health emergency of continental security (PHECS), highlighting the disproportionate impact on the African continent.[2]

Mpox, once considered a rare and geographically limited disease, has rapidly transformed into a global health threat. The 2022 outbreak, which spread to over 40 countries, revealed the virus's potential for widespread transmission across diverse populations. By 2024, the situation in Africa, particularly in the Democratic Republic of Congo (DRC), had reached critical levels, with the DRC accounting for over 96% of all cases on the continent. The emergence of clade 1b in the DRC, which has a much higher mortality rate than clade 2 that was responsible for the 2022 outbreak, is a significant global concern.[3]

The rapid spread in resource-limited settings is exacerbated by several factors:

1. Inadequate healthcare infrastructure and manpower
2. Limited access to diagnostic tools
3. Insufficient public health resources like vaccinations
4. Compounding social determinants of health, including high rates of malnutrition and HIV co-infection

While Polymerase Chain Reaction (PCR) testing remains the gold standard for Mpox diagnosis, it presents significant barriers in low-resource settings due to high costs, need for specialized equipment, and requirement for trained personnel.[4,5] Consequently, the true burden of Mpox in these regions is likely underestimated, leading to suboptimal allocation of vaccines and treatments.[4] The pressing need for innovative, scalable, and rapid screening solutions in resource-constrained environments has never been more apparent. Artificial Intelligence (AI) and Machine Learning (ML) technologies have shown promise in enhancing infectious disease screening. Recent Systematic Review have demonstrated that deep learning models, particularly convolutional neural networks (CNNs), can achieve high levels of accuracy, sensitivity, and specificity in Mpox detection, often matching traditional screening and diagnosing method.[6,7,8]

However, a critical limitation of existing AI tools is their reliance on cloud-based systems and internet connectivity, rendering them impractical in many remote or under-resourced areas.[9,10] This study addresses this gap by introducing an AI-driven, on-device screening tool specifically designed for offline use in low-resource settings. By leveraging deep learning models optimized for mobile platforms, our tool aims to provide accurate, real-time Mpox screening without the need for internet connectivity or sophisticated laboratory infrastructure.

This research not only presents a potential solution to the immediate Mpox crisis but also offers a model for rapid, adaptable technological responses to emerging infectious diseases in resource-limited settings. By enabling healthcare workers to efficiently screen and prioritize cases in the field, this tool has the potential to significantly improve resource allocation, guide isolation measures, and enhance the overall public health response to the ongoing Mpox emergency in Africa and beyond. It is important to note that while our tool provides highly accurate screening, it is not intended to replace definitive diagnosis but rather to serve as a crucial first-line tool for rapid triage and informed decision-making in resource-constrained environments.

## Methods

### Study Design

This study was designed as a descriptive analysis to build, validate, and test an AI-driven on-device screening tool for Mpox, specifically tailored for use in low-resource settings where internet connectivity is unreliable or unavailable. The study was conducted in accordance with the ethical guidelines approved by the Institutional Review Board (IRB) of the National University of Singapore together with, all existing patient consent and data privacy protocols implemented by the HeHealth team during data collection were strictly followed. The primary goal was to create a model that could accurately screen Mpox using anonymized real and synthetic data, thereby addressing the challenges posed by traditional diagnostic methods in remote areas.

### Data Collection

The dataset utilized in this study consisted of a balanced distribution of images across three categories: Mpox, other skin conditions, and normal skin. The images included in the "Mpox group" were either confirmed by PCR testing or clinically validated by a consultant physician. The "Other Skin Conditions" class in our dataset comprised a diverse range of dermatological presentations that could potentially be confused with Mpox lesions. This category included images of syphilis ulcers, herpes skin lesions, balanitis, eczemas, wound, scars, burn skin scars, skin cancers, skin tags, melanocytic nevi, vascular lesions, acne, cysts, follicular disorders, allergy rashes and other tumors, ensuring that our model was trained to differentiate Mpox from a wide spectrum of visually similar skin conditions commonly encountered in clinical practice. Total of 10,000 image pool were used in this study for all the different aspects of the model building including synthetic data generation. Data was reviewed by two physicians for diversity in terms of ethnicity of skin color. All the data was sourced from publicly available datasets from www.kaggle.com (Skin Disease Classification dataset and Acne Recognition Dataset), https://github.com (The Skin Condition Image Network(SCIN)) and HeHealth own data with relevant consent.

Each class included 1200 images for each category, split into training (75%), validation

(10%), and testing (15%) sets. For the Mpox group training data set included synthetic data mixed at 50:50 percentage. All the other classes only included actual patient data and specifically none of the validation and testing data contained synthetic data. Synthetic data inclusion was intended to improve the model's robustness and generalizability, we incorporated synthetic data generated through diffusion models, as previously validated in related research.[11] The synthetic data was carefully curated, clinically validated by two consultant physicians for accuracy and to reflect a diverse range of skin types and lesion appearances, ensuring that the model could generalize across various demographic groups.

All the images which were used in this research were anonymized to protect patient privacy. Ethical approvals and informed consents were obtained as per the IRB guidelines. The synthetic images were generated using a text-to-image diffusion model fine-tuned with the DreamBooth technique, which enabled the creation of high-fidelity images based on a minimal set of clinically validated Mpox lesion images.

**Technology Selection**

After evaluating multiple AI architectures, YOLOv8n (You Only Look Once, version 8 nano) was selected as the final model for deployment due to its low computational requirements and high accuracy.[12] YOLOv8n was selected over MobileNetV2 and Vision Transformers due to its optimal balance of speed and accuracy, which makes it particularly well-suited for deployment in resource-constrained environments.

 Performance comparison is shown in the table below.

| Parameter | YOLOv8n | MobileNetV2 |
| --- | --- | --- |
| Number of Parameters | 1.4 Million | 3.4 Million |
| Model Weight Size | 5 MB | 35 MB |
| Epochs | 200 (Stopped early at 115) | 200 (Stopped early at 90) |
| Image Size (imgsz) | 224x224 pixels | 224x224 pixels |
| Patience for Stopping | 50 epochs | 50 epochs |
| Dropout Rate | 20% | 20% |
| Overall Accuracy | 96% | 92% |
| Processing Power | Low | Moderate |

Table 1: Model comparison

YOLOv8n operates on PyTorch 2.3.1 and was selected for its ability to run efficiently on mobile devices without requiring internet connectivity. The model's configuration included 73 layers and 1.44 million parameters, with a model weight size of 5 MB, ensuring that it could be easily deployed on edge devices.

**Training**

The model was trained using a dataset of 2700 images (900 per category). Training was conducted over 200 epochs, with early stopping implemented at 115 epochs to prevent overfitting. Preprocessing techniques such as image flipping, rotation, and adjustments to saturation and brightness were applied to enhance the model's robustness and generalization.

The training process utilized the Adam optimizer with a learning rate of 1e-4, and a dropout rate of 20% was applied to prevent overfitting. The model was trained with an image size of 224x224 pixels, which is standard for YOLO-based models, ensuring compatibility with mobile device screens. We have included the training and validation loss plots to demonstrate that the model is not overfitting, as well as Top-1 and Top-5 accuracy plots to illustrate the model's prediction accuracy.

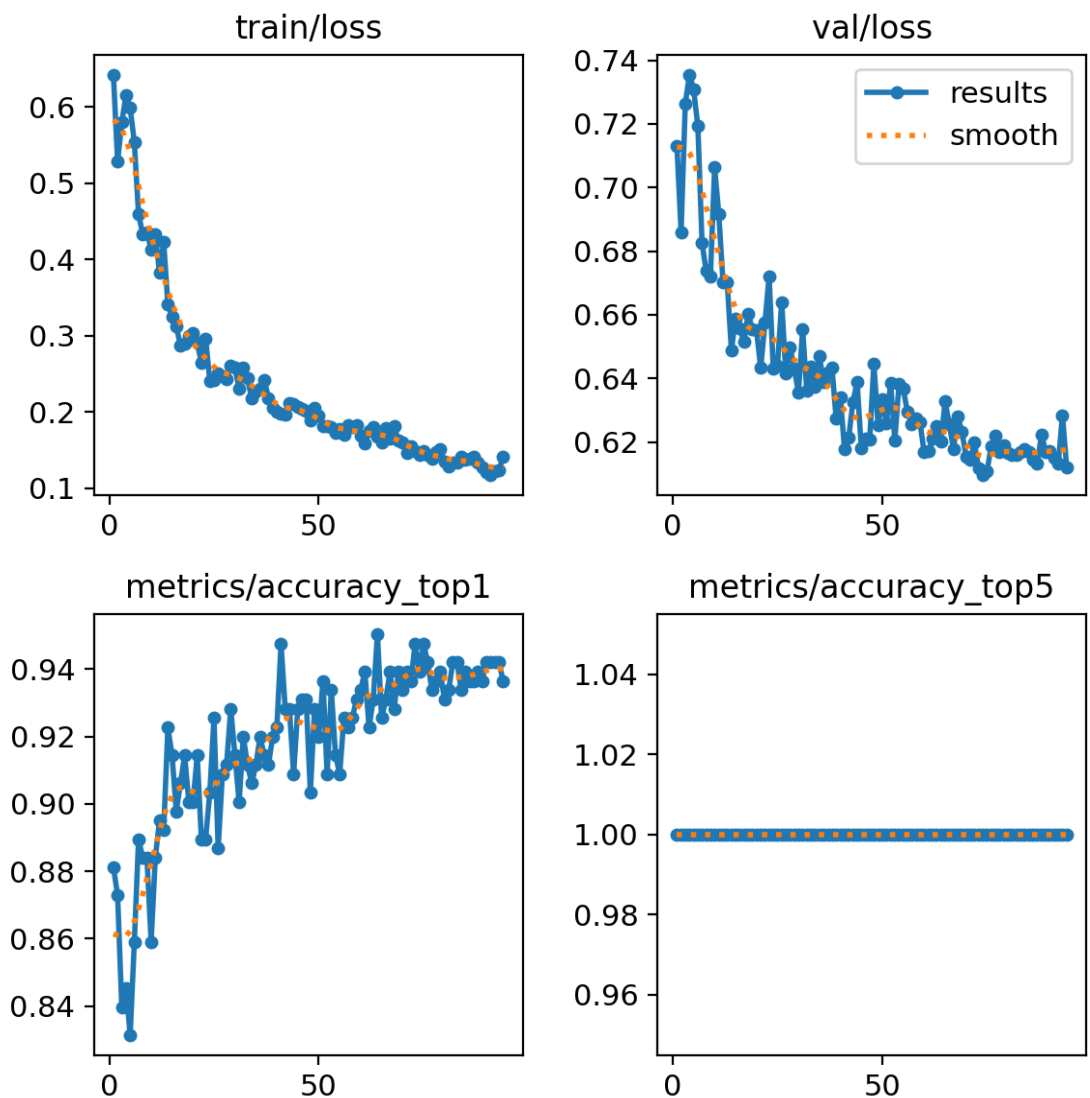

Figure 1: Training and Validation loss together with Metrics accuracy Top 1 and 5

**Initial Validation and Testing**

Validation was performed using a set of 360 images (120 per category), while testing involved 540 images (180 per category). Performance metrics such as accuracy, precision, recall, and F1-score were calculated to evaluate the model's effectiveness. Cross-validation techniques were employed to ensure the model's reliability across different subsets of the data. also, sensitivity and specificity were calculated to detect Mpox for medical use cases.

**Final validation and testing**

To further validate our model's performance and generalizability, we conducted an additional layer of external validation using a larger, independent dataset, curated by a separate team not involved in the initial project. This second Mpox dataset focused primarily on Clade 1 from Africa, sourced from open-sourced data and collaborators in Africa. It included 1,500 images, with 500 images each representing Mpox, other skin conditions, and normal skin. All 1,500 images were cross-checked for duplicity against the original dataset used for training and testing, using Python code.

**Results**

**Performance Metrics**

The YOLOv8n model demonstrated high accuracy across all categories, with an overall accuracy of 96% on the final test set. For Mpox detection, the model achieved a precision of 93%, a recall of 97%, and an F1-score of 95%. Overall accuracy of 96% demonstrated. The performance was consistent across other categories, with similar metrics reported for non-Mpox conditions and normal skin.

The performance metrics for the main categories in this final testing are as follows:

| Category | Precision (95% CI) | Recall (95% CI) | F1-score (95% CI) |
|---|---|---|---|
| Mpox | 93.1% (90.6-95.1) | 96.8% (94.9-98.1) | 94.9% (93.1-96.4) |
| Other Skin Conditions | 96.9% (94.9-98.3) | 94.4% (92.1-96.2) | 95.6% (93.9-97.0) |
| Normal | 97.8% (96.0-98.9) | 96.4% (94.5-97.8) | 97.1% (95.6-98.2) |

Table 2: Final Testing: Precision, Recall and F1 score for 3 classes

In medical terms the model achieved Sensitivity of 97% (95% CI: 95.1% - 98.3%) and Specificity of 96% (95% CI: 94.6% - 97.1%) to detect Mpox from other classes.

Results were similar to initial testing with a smaller dataset. Full results are in the table below.

| Category | Precision (95% CI) | Recall (95% CI) | F1-score (95% CI) |
|---|---|---|---|
| Mpox | 94.0% (89.5-96.8) | 93.0% (88.3-96.0) | 94.0% (90.3-96.5) |
| Other Skin Conditions | 93.0% (88.3-96.0) | 92.0% (87.1-95.3) | 93.0% (89.1-95.7) |
| Normal | 96.0% (92.0-98.2) | 98.0% (94.7-99.4) | 97.0% (93.9-98.6) |

Table 3: Initial Testing: Precision, Recall and F1 score for 3 classes

Initial results also showed Sensitivity: 93.0% (95% CI: 88.3% - 96.0%) and Specificity: 95.0% (95% CI: 92.1% - 96.9%) Both stages of testing results are also presented in a confusion matrix for the clear interpretation below.

Final testing results confusion matrix    Initial testing results confusion matrix

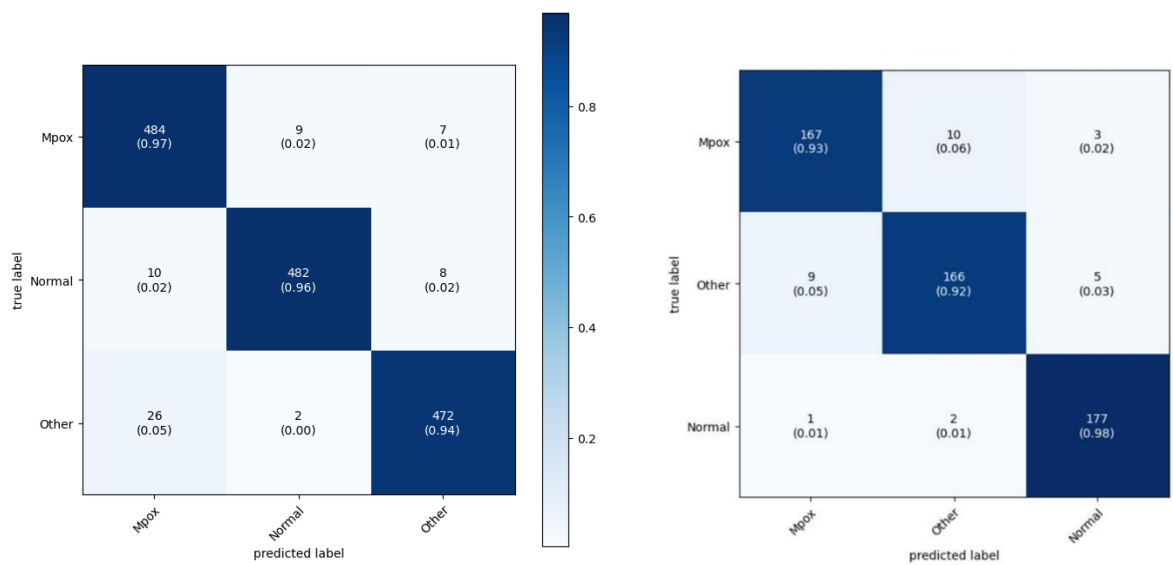

Figure 2: Final and Initial testing results confusion matrix

These results demonstrate the model's ability to maintain high performance on a larger, more diverse dataset, reinforcing its potential for real-world application.

**Operational Benefits**

The YOLOv8n model's lightweight design and high accuracy make it ideal for offline, on-device use in resource-limited areas, enabling rapid screening without internet connectivity and improving resource allocation. Field testing confirmed its user-friendliness and significant productivity boost for HCWs with minimal training.

The use of synthetic data enhanced the model's robustness, ensuring effective generalization across diverse skin types and lesion appearances. The YOLOv8n model provides an efficient Mpox screening solution for low-resource settings and can potentially be adapted for other infectious diseases.

**Limitations**

While this study provides promising results, several limitations should be acknowledged. First, the dataset used, despite being balanced and diverse, may not fully capture the global variability in Mpox presentations, which could affect the model's generalizability across different populations. Second, the reliance on synthetic data, although it has proven effective in enhancing the model's performance, may introduce biases that are not fully representative of real-world clinical scenarios. Finally, while the model has shown strong performance in initial testing, further external validation in varied healthcare settings is essential to confirm its robustness and reliability. These limitations underscore the need for ongoing research and testing to ensure the tool's efficacy in diverse, real-world environments.

**Discussion**

**Validity of the results**

A key strength of our study is the two-stage validation process. After initial testing on 540 images, we performed a larger external validation on 1,500 independent images. This rigorous approach provides strong evidence for the model's generalizability and robustness in diverse datasets, enhancing confidence in its potential real-world performance.

**Interpretation of Results**

Our AI-driven, on-device Mpox screening tool has shown to be highly effective, particularly in low-resource settings where traditional diagnostic methods like PCR are limited.[13] The tool's offline functionality and high accuracy make it a valuable asset for public health management. At the individual level, it allows for quick isolation and treatment prioritization, reducing forward transmission risk. On a broader scale at population level, it enhances surveillance, enabling more comprehensive screening, and to identify more positive cases, which is crucial for controlling outbreaks and preventing deaths. Financially, the tool offers a better return on investment by minimizing the need for costly PCR tests, allowing more efficient resource allocation.

In an ideal setting, AI screening should be used alongside PCR testing in the Mpox response to achieve multiple goals. AI screening focuses on identifying and isolating more positive cases to prevent further spread, while PCR testing validates the AI results and conducts detailed genetic analysis for viral mutations.

**Comparison with Existing Tools**

Compared to PCR, our AI tool offers significant advantages, particularly in speed, ease of use, and scalability. While PCR remains the gold standard, its requirements for specialized equipment, trained personnel and cost make it less practical in many settings.[9] Our tool, by contrast, provides rapid results on mobile devices without needing internet access. This will enable HCWs to identify more positive cases with similar resources compared to PCR testing and better return of investment for money.

**Implications for Public Health**

The deployment of this AI tool could impact public health significantly, potentially influencing WHO guidelines and setting a new standard for diagnostics in low-resource settings. AI tool can improve the accuracy of "probable case" definition, and it might be able to use it as surrogate for "confirmed case" definition in resource poor settings. Its flexibility means it could be adapted for other diseases, offering a scalable solution for future outbreaks, which is critical as global health challenges evolve.

**Open-Source Model**

By adopting an open-source approach, we aim to encourage collaboration with local African AI communities, accelerating innovation and ensuring the tool's ongoing relevance. This model supports transparency and allows for cultural and contextual adaptation, building local capacity and ensuring sustained efforts against Mpox and other diseases.

**Future Directions**

Future research should focus on validating the tool in diverse healthcare settings to ensure robustness. Expanding its application to other infectious diseases could enhance global disease surveillance. Integrating the tool into broader health systems, like personal mobile apps, electronic health records, will be crucial for maximizing its impact. Continuous updates will be necessary to keep the tool effective as AI technology and health challenges evolve.

**Conclusion**

**Summary of Findings**

This study demonstrates the effectiveness of our AI-driven, on-device Mpox screening tool, particularly in low-resource settings where traditional diagnostics like PCR are not feasible. The tool's high accuracy, ability to function offline, and strong performance across diverse demographic groups underscore its potential as a critical asset in the global fight against Mpox. By utilizing synthetic data alongside real-world images, we have developed a model that is both robust and adaptable, making it a reliable solution for rapid Mpox detection and management.

**Call to Action**

The success of this AI-driven tool highlights the urgent need for the global adoption of similar technologies, especially in regions where healthcare resources are limited. As public health challenges continue to evolve, it is essential to embrace innovation and collaboration to enhance disease surveillance and response. We urge global health organizations, governments, and technology developers to invest in and support the deployment of AI-powered, offline screening tools, not only for Mpox but also for other emerging health threats. By doing so, we can significantly improve public health outcomes and build more resilient healthcare systems worldwide.